\documentstyle[12pt,titlepage]{article}

\renewcommand{\thesection}{\arabic{section}}
\renewcommand{\theequation}{\thesection.\arabic{equation}}
\def\bl{\Biggl\{}
\def\br{\Biggr\}}
\def\bpl{\Biggl(}
\def\bpr{\Biggr)}

\def\L{{\cal L}}

\def\d{\delta}
\def\D{\Delta}
\def\e{\epsilon}

\def\M{{\cal M}}
\def\O{{\cal O}}

\def\gbig {\hbox{\Large\it g}}
\def\J{{\cal J}}
\def\k{\kappa}
\def\l{\lambda}

\def\so{\emptyset}

\def\z{\zeta}

\def\s{\sigma}

\def\z{\zeta}

\def\vphi{\varphi}
\def\w{\omega}

\def\hf{\frac{1}{2}}
\def\der{\partial}

\def\bq{\begin{equation}}
\def\eq{\end{equation}}
\def\brr{\begin{eqnarray}}
\def\err{\end{eqnarray}}
\def\ba{\left(\begin{array}}
\def\ea{\end{array}\right)}

\def\lapp{\stackrel{<}{\sim}}

\def\ba{\left(\begin{array}}
\def\ea{\end{array}\right)}
\input epsf.tex
\epsfverbosetrue
\begin{document}
\pagestyle{empty}
\begin{flushright} CERN-TH.7356/94, UPR-620T \end{flushright}
\begin{large}
\begin{center}{\bf THE STRENGTH OF NON-PERTURBATIVE EFFECTS }\end{center}
\begin{center}{\bf IN MATRIX MODELS}\end{center}
\begin{center}{\bf AND STRING EFFECTIVE LAGRANGIANS}\end{center}

\end{large}
\begin{center} RAM BRUSTEIN  \end{center}
\vspace{-.3in}
\begin{center} Theory Division, CERN \end{center}
\vspace{-.3in}
\begin{center} CH-1211 Geneva 23, Switzerland  \end{center}

\begin{center} MICHAEL FAUX$^{*)}$ and BURT A. OVRUT$^{*,**)}$
\end{center}
\vspace{-.3in}
\begin{center} Department of Physics \end{center}
\vspace{-.3in}
\begin{center} University of Pennsylvania \end{center}
\vspace{-.3in}
\begin{center} Philadelphia, Pa 19104-6396 \end{center}
\begin{center} ABSTRACT \end{center}

We present a summary of the results  of  an explicit  calculation of the
strength of non-perturbative interactions in matrix models and string effective
Lagrangians. These interactions  are induced  by single eigenvalue instantons
in the  $d=1$ bosonic matrix model. A well defined approximation scheme is used
to obtain
induced operators whose exact form we exhibit. We briefly discuss
the possibility  that similar instantons in a  supersymmetric version of the
theory may break supersymmetry dynamically.

\vspace*{1.8cm}
\noindent
\rule[.1in]{16.5cm}{.002in}\\
\noindent
$^{*)}$ Work supported in part by DOE under Contract No.
DOE-AC02-76-ERO-3071.\\
$^{**)}$ Talk presented at the SUSY94 Workshop,
May 14-171994, Ann Arbor, Michigan.

\vspace*{0.5cm}

\begin{flushleft} CERN-TH.7356/94 \\
July 1994
\end{flushleft}
\vfill\eject

\setcounter{page}{1}
\pagestyle{plain}

\renewcommand{\thesection}{\arabic{section}}
\renewcommand{\theequation}{\arabic{equation}}
\setcounter{equation}{0}

Recently, it has been shown that matrix models \cite{mm} allow the
construction of space-time Lagrangians valid to all orders in the
string coupling parameter,
at least for noncritical strings propagating in $d=2$ dimensions.
These Lagrangians are derived using the techniques of collective field theory
\cite{collft,jevrev}.   All order  Lagrangians have been constructed, using
these techniques, for both the $d=1$ bosonic matrix model \cite{bo} and also
for the $d=1, {\cal N}=2$ supersymmetric matrix model \cite{sbfo}. There are
two remarkable features of these constructions.  First, when interactions are
included to all orders, the induced coupling blows up at finite points in space
and delineates a zone of strong coupling.  This is to be contrasted with the
lowest order theory, where
the coupling only diverges at spatial infinity.  Secondly, since these
all-order Lagrangians are derived from matrix models, they contain additional
non-perturbative information which is directly
accessible and computable.   The existence of these new non-perturbative
aspects of the theory relies on the observation
that the matrix models contain two distinct sectors.  The first of
these is the so-called continuous sector, which consists of a
continuous distribution of matrix eigenvalues. The second sector consists of
discrete eigenvalues, which are distinguishable from the continuum eigenvalues.
 The classical configurations
of the matrix model include time-dependent instanton solutions in which the
discrete eigenvalues tunnel between two continuous eigenvalue sectors.  We
perform an explicit  calculation of the leading order effects of such single
eigenvalue instantons on the effective theory derived from a $d=1$ bosonic
matrix model.
These consists of a set of induced operators, whose exact form we
compute and exhibit. The results presented here are a summary
of the results contained in \cite{bfo}. All  calculations are presented in
painful details there.

We conjecture that, in the supersymmetric case, the same instantons described
in this talk, and their
associated bosonic and fermionic zero modes, provide a mechanism
for supersymmetry breaking in the associated $d=2$ effective
superstring theory. It is presumed that the discrete nature of the
single eigenvalues allows a novel circumvention of some no-go theorems, based
on Witten's index, relevant to non-perturbative dynamical supersymmetry
breaking in $d>1$ dimensions.  The present calculation is  a necessary
preliminary  ingredient to the explicit calculation of this effect, which we
are pursuing at these very moments and hope to report on soon \cite{nbfo}.
Non-perturbative effects due to single eigenvalue instantons
and their implications were also discussed elsewhere
\cite{shenker,dabh,leemend,dmw}. Recently, an interesting  complementary
approach was suggested \cite{joep}.

A $d=1$ bosonic matrix model has a time-dependent $N\times N$ Hermitian
matrix, $M(t)$, as its fundamental variable.  Its dynamics are
described by the
Lagrangian
\bq L(\dot{M},M)=\hf Tr \dot{M}^2-V(M).
 \label{lagm}\eq
The potential is taken to be polynomial,
\bq V(M)=\sum_{n=0}^\infty a_n Tr M^n,
 \label{v1def}\eq
As $N\rightarrow\infty$, if the $a_n$ are
tuned simultaneously and appropriately, the associated partition function
describes an
ensemble of oriented two-dimensional
Riemann surfaces, including contributions at all genus.
It is argued that, in this limit,
the model describes a string propagating in two
space-time dimensions. In the large $N$ limit, the
potential may be written as
\bq V(M)=Tr(N{V}_0\cdot {\bf 1}-\hf\w^2M^2),
 \label{vdef}\eq
where ${\bf 1}$ is the
$N\times N$ unit matrix.  The parameters ${V}_0$ and $\w$ each
have mass dimension one, and are arbitrary.  In (\ref{vdef}) the
scaling behavior of the coefficients has been made explicit.
The Lagrangian, (\ref{lagm}), is invariant under the
global $U(N)$ transformation $M\rightarrow{\cal U}^\dagger M{\cal U}$, where
${\cal U}$ is an arbitrary $N\times N$ unitary matrix.  The set of states which
do not transform under ${\cal U}$ comprise the
$U(N)$-singlet sector of the quantized theory.
It can be shown that the physics of this singlet sector is described
equivalently by a theory involving only the $N$ eigenvalues, $\l_i(t)$,
of the
matrix $M(t)$ with the following Lagrangian,
\bq L[\l]=\sum_{i=1}^N\{\hf\dot{\l}_i^2-({V}_0-\hf\w^2\l_i^2)
    -\hf\sum_{j\ne i}\frac{1}{(\l_i-\l_j)^2}\}.
 \label{lageig}\eq
The eigenvalues are first restricted to lie in the interval
$-\frac{L}{2}\le \l_i\le\frac{L}{2}$ for any $i$.
When we take the limit
$N\rightarrow\infty$, we will simultaneously take
$L\rightarrow\infty$.  In this limit, over a given range, $l$,
to be made explicit below,
there exist two possibilities.
If $n$ represents the number of eigenvalues within this range,
then the average density is given by $\rho=n/l$.  In the limit
$N\rightarrow\infty, L\rightarrow\infty$, $\rho$ can remain small,
and the eigenvalues populate the region sparsely.
We refer to this situation
as a ``low density" or ``discrete" distribution of eigenvalues over the
region
$l$.
In the second case, $\rho$ becomes very
large, and the eigenvalues
populate the region densely. In this case, the eigenvalues can
be aggregated into a ``collective field" which describes their  collective
motion.  We refer to this second case as a ``high density"
or``continuous"
distribution
of eigenvalues.
We begin by studying the continuous case.

We introduce a continuous
real parameter, $x$, constrained to lie in the interval
$-\frac{L}{2}\le x\le\frac{L}{2}$, and over this line segment define a
collective
field,
\bq \der_x\vphi(x,t)=\sum_{i=1}^N\d(x-\l_i(t)).
 \label{pdef}\eq
It follows from (\ref{pdef}) that
\bq \int_{x_0}^{x_0+l}dx\der_x\vphi(x,t)=n,
 \label{rel}\eq
where $n$ is the number of eigenvalues in the range $l$.
Thus, $\vphi'=\der_x\vphi$ is
the eigenvalue density.  In the range $l$,
$\vphi'$ has $n$ degrees of freedom.  Provided that
$n/l\rightarrow\infty$ as $N\rightarrow\infty, L\rightarrow\infty$,
the average density of eigenvalues then becomes infinite, and,
modulo some technical subtleties irrelevant to this discussion,
the field $\vphi$ becomes an unconstrained, ordinary
two dimensional field.
In effect,
$\vphi'$ ceases to be a sum over delta
functions and becomes a continuous eigenvalue density.
It can be shown, in this case,  that
the eigenvalue Lagrangian, (\ref{lageig}), may be rewritten in terms of
the
collective field as follows,
\bq L[\vphi]=\int
dx\{\frac{\dot{\vphi}^2}{2\vphi'}-\frac{\pi^2}{6}\vphi^{'3}
    -(V_0-\frac{\w^2}{2}x^2)\vphi'\}.
 \label{lagcoll}\eq
The associated action is given by $S[\vphi]=\int dt L[\vphi]$.
This expression describes the physics over all ranges of $x$ where the
eigenvalue density is large.  The limits on the $\int dx$ integral
are set accordingly. Since our interest is in the quantum theory, henceforth we
will consider only the Euclidean version of the action, which is given by
\bq S_E[\vphi]=\int dx dt \{\frac{\dot{\vphi}^2}{2\vphi'}
    +\frac{\pi^2}{6}\vphi^{'3}
    +(V_0-\frac{\w^2}{2}x^2)\vphi'\}.
 \label{se}\eq
The
equation of motion, obtained by varying
(\ref{se}) is
\bq \der_t(\frac{\dot{\vphi}}{\vphi'})
    -\hf\der_x\bl\frac{\dot{\vphi}^2}{\vphi^{'2}}+\pi^2\vphi^{'2}
    -\w^2x^2\br=0.
 \label{momo}\eq
The static solution is obtained by taking $\dot{\vphi}=0$, so that
(\ref{momo})
reduces to
\bq \der_x\bl\pi^2\vphi^{'2}-\w^2x^2\br=0.
 \label{eqmo}\eq
The solution to this equation is the following,
\bq \tilde{\vphi}_0^{'}(x)=\frac{\w}{\pi}\sqrt{x^2-A^2},
 \label{vback}\eq
where $A^2$ is a positive constant. Additional analysis reveals that \bq
A^2=2V_0/\w^2.
 \label{aak}\eq

Since $\vphi'$ is now a continuous density of eigenvalues, we may use
(\ref{rel}) to determine the approximate location of the
first eigenvalues in the continuum; that is, those
two eigenvalues closest to $x=\pm A$.
We focus on the region $x\ge A$.
There is an identical discussion regarding the opposite region,
$x\le-A$.
Given (\ref{vback}), the first eigenvalue must live
somewhere in the region $A\le x\le A+\e_x$, where $\e_x$ is determined
by
the following relation,
\brr 1 &=& \frac{\w}{\pi}\int_A^{A+\e_x}dx\sqrt{x^2-A^2} \nonumber \\
       &=& \frac{\w A^2}{2\pi}\bl\frac{x}{A}\sqrt{(\frac{x}{A})^2-1}-
       \ln(\frac{x}{A}+\sqrt{(\frac{x}{A})^2-1})\br
       \Biggl{|}_{x=A}^{x=A+\e_x}.
 \label{ky}\err
We make the important assumption that $\e_x<<A$.  After some algebra,
Eq.(\ref{ky}) then becomes
\bq \hf(\frac{3\pi}{\w A^2})^{2/3}=\frac{\e_x}{A}
    +{\cal O}\bpl(\frac{\e_x}{A})^2\bpr.
 \eq
For consistency, this requires that $(\w A^2)^{-1}<<1$.
This small dimensionless number will be central to much of the ensuing
analysis, so we give it a special name,
\bq g=\frac{1}{\w A^2}<<1.
 \label{gdef}\eq
It is clear that the first eigenvalue does not live
precisely at the value $x=A$.  This distinction will prove a necessary
and important regulator on quantities which we will encounter.  For
definiteness, we assume henceforth that the first eigenvalue in the
static
continuum has a value $x=A+\e_x$, where
\bq \e_x=\hf(3\pi g)^{2/3}A
 \label{edef}\eq
and $g$ is a small, dimensionless number, which,
in the present context, parameterizes the width of the discrete region as well
as our ignorance
regarding the ``graininess" of eigenvalues near the edge of the
continuous distribution, when we adopt a collective
field point of view.
We now turn our attention to the region $|x|\le A$.
We assume, in addition to a continuum of eigenvalues
$\l_i$ for $i=1$ to $N$, that there exists an additional
discrete eigenvalue, which we
denote $\l_0$.   There are then $N+1$
total eigenvalues, and the Euclidean version of Lagrangian
(\ref{lageig}) now reads
\brr L_E = \sum_{i=0}^N\{\hf\dot{\l}_i^2+({V}_0-\hf\w^2\l_i^2)
    +\hf\sum_{j\ne i}\frac{1}{(\l_i-\l_j)^2}\}.
 \err
Note that the index $i$ now runs over the $N+1$ values from $0$ to $N$.
What do we mean by a discrete eigenvalue?  The separation of the continuum
eigenvalues nearest to
$\pm A$ is of order $\e_x$.  As long as $-A\le\l_0\le A$, and
\bq A-|\l_0|>>\e_x,
 \label{condo}\eq
the eigenvalue $\l_0$ is truly distinct from the continuum and,
hence, discrete.
Assuming that $\l_0$ satisfies (\ref{condo}),
it is useful to rewrite this Lagrangian by separating the $\l_0$
contribution
from the contribution due to the continuum eigenvalues, as follows,
\brr L_E &=& \hf\dot{\l}_0^2+(V_0-\hf\w^2\l_0^2)
     +\sum_{i\ne 0}\frac{1}{(\l_0-\l_i)^2} \nonumber \\
     &+&\sum_{i=1}^N\{\hf\dot{\l}_i^2+({V}_0-\hf\w^2\l_i^2)
     +\hf\sum_{j\ne i}\frac{1}{(\l_i-\l_j)^2}\}.
\err
As above, we may now rewrite this expression using the definition
(\ref{pdef}).  We thus obtain
\brr L_E[\l_0;\vphi] &=& \hf\dot{\l}_0^2+\hf\w^2(A^2-\l_0^2)
     +\int dx\frac{\vphi'}{(x-\l_0)^2} \nonumber \\
     &+& \int dx\{\frac{\dot{\vphi}^2}{2\vphi'}
     +\frac{\pi^2}{6}\vphi^{'3}
     +\hf\w^2(A^2-x^2)\vphi'\}.
 \label{hybrid}\err
The third term in this expression represents the mutual
interaction of the discrete eigenvalue with the continuum eigenvalues,
which are collectively described using the field $\vphi$.
We obtain the Euclidean equations of motion for $\l_0$ and for $\vphi$
by variation of (\ref{hybrid}).  Respectively, these are found to be
\bq \ddot{\l}_0+\w^2\l_0+\int dx\frac{\vphi'}{(\l_0-x)^3}=0
 \label{mo1}\eq
\bq \der_t(\frac{\dot{\vphi}}{\vphi'})
    -\hf\der_x\bl\frac{\dot{\vphi}^2}{\vphi^{'2}}+\pi^2\vphi^{'2}
    -\w^2x^2+\frac{2}{(\l_0-x)^2}\br=0.
 \label{mo2}\eq
We consider first the $\vphi$ equation.
It is possible  to show, even in the presence of a nontrivial,
but discrete, $\l_0(t)$,
that the static background, $\tilde{\vphi}_0^{'}$, derived above
is still a valid solution to leading order in $\e_x$.

Next, we turn our attention to the $\l_0$ equation, (\ref{mo1}).  This
is
the Euclidean equation of motion,
\bq \ddot{\l}_0-V_{eff}^{'}(\l_0)=0,
 \label{moe}\eq
where
\bq V_{eff}(\l_0)=\frac{\w}{2g}\bl
    -(\frac{\l_0}{A})^2+4g\frac{(\l_0/A)}{\sqrt{1-(\l_0/A)^2}}
    \tan^{-1}(\frac{(\l_0/A)}{\sqrt{1-(\l_0/A)^2}})\br.
 \label{vvv}\eq
The effect of the second term in
(\ref{vvv}),
is to
turn the potential over near $\l_0=\pm A$, where it adds infinite
confining walls.  The eigenvalue, $\l_0$ can be treated as
discrete, and $V_{eff}(\l_0)$ is well defined, for $\l_0$ sufficiently
far from $\pm A$.  When $\l_0$ approaches $\pm A$ to within order $\e_x$
it is absorbed into the continuum, and disappears as a discrete entity. Of
course, this process can be reversed.  It is possible for the first eigenvalue
of the continuum to ``leak" out and become a discrete eigenvalue $\l_0$.  We
will return to such processes below.

This being said, we would like to find both static and time-dependent
solutions for the Euclidean $\l_0$ equation of motion (\ref{moe}).    In the
small $g$ limit we can
replace (\ref{moe}) by
\brr \ddot{\l}_0+\w^2\l_0 &=& 0
     \hspace{.3in}; \hspace{.1in} -A<\l_0<A \nonumber \\
     \ddot{\l}_0 &=& 0
     \hspace{.3in}; \hspace{.1in} \l_0=\pm A.
 \label{keq} \err
We also impose the following boundary conditions,
$\l_0(t\rightarrow -\infty)= \pm A$ and, independently,
$\l_0(t\rightarrow +\infty)= \pm A$.
There are two static solutions to (\ref{keq}) which satisfy this
boundary
condition,
\bq {\widehat\l}_{0\pm}=\pm A.
 \eq
A simple time-dependent solution is given
by
\bq {\widehat\l}_0^{(+)}(t;t_1)=\left\{
   \begin{array}{ll}
    -A & ;\hspace{.2in} t<t_1-\frac{\pi}{2\w} \\
    +A\sin\w(t-t_1) & ;\hspace{.2in}
    t_1-\frac{\pi}{2\w}\le t\le t_1+\frac{\pi}{2\w} \\
    +A & ;\hspace{.2in} t>t_1+\frac{\pi}{2\w} \end{array}\right.,
 \label{lop1}\eq
where $t_1$ is arbitrary.
The solution (\ref{lop1})
describes an eigenvalue which rolls (tunnels) from $-A$ to $+A$ over
a time interval of duration $\frac{\pi}{\w}$, centered at an arbitrary
time $t_1$.
We refer to this solution as a ``kink".  Its mirror image
is also a valid solution,
\bq {\widehat\l}_0^{(-)}(t;t_1)=\left\{
   \begin{array}{ll}
    +A & ;\hspace{.2in} t<t_1-\frac{\pi}{2\w} \\
    -A\sin\w(t-t_1) & ;\hspace{.2in}
    t_1-\frac{\pi}{2\w}\le t\le t_1+\frac{\pi}{2\w} \\
    -A & ;\hspace{.2in} t>t_1+\frac{\pi}{2\w} \end{array}\right.,
 \label{lop2}\eq
It
describes an eigenvalue which rolls
from $+A$ to $-A$.
It is referred
to as an ``anti-kink".

Taking into account the fact that, when at $\pm A$, the discrete eigenvalue
gets reabsorbed in the continuum,   we may rewrite the
kink and antikink solutions as follows,
\bq {\l}_0^{(\pm)}=\pm A\sin\w(t-t_1)
    \hspace{.2in} ; \hspace{.1in}
    t_1-\frac{\pi}{2\w}\le t\le t_1+\frac{\pi}{2\w},
 \label{klink}\eq

There exist more general solutions than those which we have already discussed,
in which the identity of $\l_0$ is  more complex.  It is possible, for example,
that a kink, which ends with eigenvalue $\l_0$ attaching to the continuum at
$+A$, could be
followed, at some  later time, by an antikink, in which the eigenvalue $\l_0$
separates from the continuum at $+A$, rolls to $-A$ and then reattaches there.
Such a kink-antikink sequence, which we denote $\l_0^{(+-)}$,
would satisfy the Euclidean equation of motion,  (\ref{keq}).  It is also
possible, however, that a kink, which ends with   the eigenvalue $\l_0$
attaching to the continuum at $+A$, could be followed, at some  later time, by
another kink in which a
different eigenvalue detaches from the continuum at $-A$, traverses the region
between $-A$ and $+A$, and then reattaches to the continuum at $+A$ immediately
next to the eigenvalue involved in the first kink. This kink-kink sequence,
which we denote $\l_0^{(++)}$, also satisfies (\ref{keq}).  There are thus
$2^2=4$ solutions which involve two
distinct
kinks,
\brr {{\l}}_0^{(++)}
   &=&
    \left\{\begin{array}{ll}
    +A\sin\w(t-t_1) &
    ;\hspace{.2in} t_1-\frac{\pi}{2\w}\le t\le t_1+\frac{\pi}{2\w} \\
    +A\sin\w(t-t_2) &
    ;\hspace{.2in} t_2-\frac{\pi}{2\w}\le t\le t_2+\frac{\pi}{2\w}
    \end{array}\right. \nonumber \\
   {\l}_0^{(+-)}
   &=&
    \left\{\begin{array}{ll}
    +A\sin\w(t-t_1) &
    ;\hspace{.2in} t_1-\frac{\pi}{2\w}\le t\le t_1+\frac{\pi}{2\w} \\
    -A\sin\w(t-t_2) &
    ;\hspace{.2in} t_2-\frac{\pi}{2\w}\le t\le t_2+\frac{\pi}{2\w}
    \end{array}\right. \nonumber \\
   {\l}_0^{(-+)}
   &=&
    \left\{\begin{array}{ll}
    -A\sin\w(t-t_1) &
    ;\hspace{.2in} t_1-\frac{\pi}{2\w}\le t\le t_1+\frac{\pi}{2\w} \\
    +A\sin\w(t-t_2) &
    ;\hspace{.2in} t_2-\frac{\pi}{2\w}\le t\le t_2+\frac{\pi}{2\w}
    \end{array}\right. \nonumber \\
   {\l}_0^{(--)}
   &=&
    \left\{\begin{array}{ll}
    -A\sin\w(t-t_1) &
    ;\hspace{.2in} t_1-\frac{\pi}{2\w}\le t\le t_1+\frac{\pi}{2\w} \\
    -A\sin\w(t-t_2) &
    ;\hspace{.2in} t_2-\frac{\pi}{2\w}\le t\le t_2+\frac{\pi}{2\w}
    \end{array}\right.
 \label{lopp}\err
In all four cases
$t_2\ge t_1+\frac{\pi}{\w}$, but both $t_1$ and $t_2$ are otherwise
arbitrary.
An arbitrary solution consists of $q$ events which are randomly
distributed between kinks and antikinks, where $0\le q<\infty$.
For a given $q$ there are $2^q$ distinct instanton configurations.
Generically, we denote the $2^q$ instantons as $\l_0^{(q)}$.  There are $q$
zero modes associated with each $\l_0^{(q)}$.  These correspond to the
arbitrary times
$t_1,...,t_q$,
where $t_q\ge t_{q-1}
\cdots\ge t_1$, when the kinks or antikinks occur.
We ignore all cases where several eigenvalues are
simultaneously discrete, since the effect of these solutions is
negligible.

The partition function associated with the theory discussed above can be
written as a sum over different instanton sectors,
\bq Z=\sum_{q=0}^\infty Z_q
 \label{kashi}\eq
where, schematically,
\bq Z_q=\int[d\vphi]\int[d\l_o]_qe^{-S[\l_0;{\vphi}]}.
 \label{schem}\eq
In this expression
the symbol $[d\l_0]_q$ indicates that $\l_0$ is expanded around
$\l_0^{(q)}$.
For notational convenience we have suppressed a subscript $E$
on the action, but it is assumed throughout this section that we
are in euclidean space. We proceed to define equation
(\ref{schem}) in more precise terms.  First of all, remember that
$\l_0^{(q)}$
generically represents all the $2^q$ instanton solutions which each have $q$
single eigenvalue kinks-antikinks.  Therefore, more specifically,
\bq Z_q=\sum_{\{k_i\}}Z_{k_1\cdots k_q},
 \label{bukhara}\eq
where $k_i=\pm$, the summation is over all $2^q$
possible sets $\{k_1\cdots k_q\}$, and
\bq Z_{k_1\cdots k_q}=\int[d{\vphi}]\int[d\l_0]_{k_1\cdots k_q}
    e^{-S[\l_0;{\vphi}]}.
 \eq
The symbol $[d\l_0]_{k_1\cdots k_q}$ indicates that $\l_0$ is expanded
around $\l_0^{(k_1\cdots k_q)}$.  Thus,
$Z_2=Z_{++}+Z_{+-}+Z_{-+}+Z_{--}$, and so on.
After some lengthy  analysis, using a dilute gas approximation, we arrive at
the following general
result
  \brr Z &=& \int[d\vphi]e^{-S_\vphi[\vphi]}\sum_{q=0}^\infty
     \frac{1}{q!}\M^q\prod_{i=1}^q
     \int dt_i\sum_{\{k_i\}}\prod_{j=1}^q
     e^{-S_I^{(k_j)}[\vphi;t_j]} \nonumber \\
     &=& \int[d\vphi]e^{-S_\vphi[\vphi]}\sum_{q=0}^\infty
     \frac{1}{q!}
     \bl\M\int dt_1\bpl e^{-S_I^{(+)}[\vphi;t_1]}
     +e^{-S_I^{(-)}[\vphi;t_1]}\bpr\br^q.
 \err
The sum over $q$ is now an exponential, so that
\bq Z=\int[d\vphi]e^{-S_{eff}[\vphi]},
 \eq
where
\bq S_{eff}[\vphi]=S_\vphi[\vphi]+\Delta S[\vphi]
 \eq
is the effective action with the instanton effects systematically
incorporated, and
\bq \Delta S[\vphi]=\M\int dt_1
    \bl e^{-S_I^{(+)}[\vphi;t_1]}
     +e^{-S_I^{(-)}[\vphi;t_1]}\br
 \label{delta}\eq
is the associated change in the action.
The action  $S_I^{(\pm)}$ is given by
\bq
S_I^{(\pm)}[\vphi;t_j]=
\int_{t_j-\frac{\pi}{2\w}}^{t_j+\frac{\pi}{2\w}}dt
\int dx\bl\frac{\vphi'(x,t)}{(x-\l_0^{(\pm)}(t-t_j))^2}
-\frac{\vphi'(x,t)}{(x-\l_{\so}^{(\pm)}(t-t_j))^2}\br.
 \label{dali}\eq
where
\bq \l_{\so}^{(\pm)}(t;t_1)=\left\{\begin{array}{ll}
      \mp A & ;\hspace{.2in} t_1-\frac{\pi}{2\w}\le t<t_1 \\
      \pm A &  ;\hspace{.2in} t_1<t\le t_1+\frac{\pi}{2\w}
\end{array}\right..
 \label{step}\eq
The quantity $\M$ is a dimensionful parameter that sets the
basic strength for induced non-perturbative interactions
\bq \M=\w\sqrt{\frac{\pi}{2g}} e^{-\frac{\pi}{2g}}.
\label{madef}\eq

So far, we have studied the collective field theory
expressed in terms of the field $\vphi$.  By examining equation
(\ref{se}),
however, we discover that $\vphi$ does not have a canonically
normalized kinetic energy.  We also find that the collective field
Lagrangian is neither Lorentz invariant nor translation invariant.
The first of these problems is solved, in part, by expanding $\vphi$
around
the solution to the euclidean field equation $\tilde{\vphi}_0$ given in
(\ref{vback}).  Thus, we define
\bq \vphi(x,t)=\tilde{\vphi}_0(x)+\frac{1}{\sqrt{\pi}}\z(x,t).
 \eq
As discussed at length elsewhere, a canonical kinetic energy
is obtained by expressing the Lagrangian in terms of
a new spatial coordinate $\tau$ defined by the following
relation,
\bq \tau'(x)=\frac{1}{\pi}(\tilde{\vphi}_0^{'}(x))^{-1}.
 \label{tox}\eq
Note that $\tau$ has mass dimension $-1$, which is the appropriate mass
dimension for a spatial coordinate, whereas $x$ has mass dimension
$-\hf$.  Expressing the euclidean collective field action (\ref{se}) in
terms of $\z(\tau,t)$, we find, in the absence of instanton effects,
that
\bq S_\z[\z] = \int dt\int d\tau\bl
       \hf(\dot{\z}^2+\z^{'2})
       -\hf\frac{\gbig(\tau)\dot{\z}^2\z'}{1+\gbig(\tau)\z'}
       +\frac{1}{6}\gbig(\tau)\z^{'3}
       -\frac{1}{3}\frac{1}{\gbig(\tau)^2}\br,
 \label{zest} \eq
where $\gbig(\tau)$ is a space dependent coupling parameter, which we
define below, and the $\tau$ integration is over the limits
$-\infty<\tau\le \tau_0+\frac{\s}{2}$
and $\tau_0+\frac{\s}{2}\le\tau<\infty$,
where $\tau_0$ and $\s$ are independent integration constants which
arise when solving (\ref{tox}).
The reason why there are two integration
constants rather than one, given that (\ref{tox}) is a first-order
differential equation, is that we must solve (\ref{tox}) independently
over
the two separate regions $-\infty<x\le A$ and $A\le x<\infty$.
The region $-A<x<A$, where there is no continuous
collective field theory, is the low density region. In $\tau$ space,
this region is given by $\tau_0-\frac{\s}{2}<\tau<\tau_0+\frac{\s}{2}$,
so that
$\tau_0$ is the center of the low density region and $\s$ is the
width.
The coupling parameter, defined over
$-\infty<\tau\le \tau_0-\frac{\s}{2}$ and
$\tau_0+\frac{\s}{2}\le\tau<\infty$,
is given by $\gbig(\tau)=(\pi^{3/2}\tilde{\vphi}_0(x))^{-1}$, and
is found to be
\bq \gbig(\tau)=
     4\sqrt{\pi}\frac{g}{\w}\frac{\frac{1}{\k} e^{-2\w|\tau-\tau_0|}}
     {(1-\frac{1}{\k}e^{-2\w|\tau-\tau_0|})^2},
 \label{ggdef}\eq
where $\k$ is a dimensionless number,
\bq \k=\exp(-\w\s),
 \label{kdef}\eq
which relates the width, $\s$, of the low density region in $\tau$
space
to the natural length scale in the matrix model, $1/\w$.
Notice that the coupling
parameter blows up as $\tau\rightarrow\tau_0\pm\frac{\s}{2}$; that is,
at
the boundaries of the low density region.

We would now like to express the change in the effective action due to the
instanton effects, equation (\ref{delta}), in terms of the canonical variable
$\z(\tau,t)$.  Since $S_I^{(\pm)}$ is linear in
$\vphi$, it follows that
\bq S_I^{(\pm)}[\vphi;t_1]=S_I^{(\pm)}[\tilde{\vphi}_0]
    +\frac{1}{\sqrt{\pi}}S_I^{(\pm)}[\z;\tau_0,t_1].
 \label{erg}\eq
The $\tau_0$ dependence in the last term of this equation will be
made clear presently.
{}From (\ref{dali}), we find
\bq
S_I^{(\pm)}[\z;\tau_0,t_1]=\int_{t_1-\frac{\pi}{2\w}}^{t_1+
\frac{\pi}{2\w}}dt\int
d\tau\bl\frac{\z'(\tau,t)}{(x(\tau)-\l_0^{(\pm)}(t-t_1))^2}
    -\frac{\z'(\tau,t)}{(x(\tau)-\l_{\so}^{(\pm)}(t-t_1))^2}\br,
 \label{sox}\eq
where the prime now means differentiation with respect to $\tau$,
and where
\bq x(\tau)=\left\{\begin{array}{ll}
    -A\cosh\{\w(\tau-\tau_0+\s/2)\} & \hspace{.2in} ;
\tau\le\tau_0-\s/2 \\
     +A\cosh\{\w(\tau-\tau_0-\s/2)\} & \hspace{.2in} ;
\tau\ge\tau_0-\s/2
     \end{array}\right. .
 \label{xot}\eq
This last expression is found by integrating (\ref{tox}) to obtain
$\tau(x)$ and then inverting the result to obtain $x(\tau)$.
This function depends explicitly on $\tau_0$.  This explains why there
is an explicit $\tau_0$ in equations (\ref{erg}) and (\ref{sox}).
It is straightforward to compute $S_I^{(\pm)}[\tilde{\vphi}_0]$ and we find
\bq S_I^{(\pm)}[\tilde{\vphi}_0]
     = -2^{3/2}\sqrt{\frac{A}{\e_x}}
     +\ln\sqrt{\frac{A}{\e_x}}+\O(\frac{\e_x}{A}).
 \label{ploonp}\eq
As discussed above, $\e_x$ is the size
of the inter-eigenvalue separation near the edge of the continuum and
so
provides the natural regulator for expressions such as (\ref{ploonp}).
{}From (\ref{edef}) it follows that, to lowest order in $g$
\bq e^{-S_I^{(\pm)}[\tilde{\vphi}_0]}= g^{1/3}e^{\O(g^{1/3})}.
 \label{ploon}\eq
Since all $x$-space integrations are cut-off at a distance $\e_x$ from
the
edge of the low density region; that is, at $|x|=A+\e_x$, it follows
that
all $\tau$ space integrals must be cut-off as well at a value
$\e_\tau$.
Specifically, in (\ref{sox}) and in all other expressions which include a $\int
d\tau$ integration, the following is implied,
\bq \int d\tau=\int_{-\infty}^{\tau_0-\frac{\s}{2}-\e_\tau}d\tau
    +\int_{\tau_0+\frac{\s}{2}+\e_\tau}^\infty d\tau.
 \eq
The value of $\e_\tau$ is simple to obtain.  We require that
\brr x(\tau-\frac{\s}{2}-\e_\tau) &=& -A-\e_x \nonumber \\
     x(\tau+\frac{\s}{2}+\e_\tau) &=& A+\e_x.
 \err
Using (\ref{xot}) and (\ref{edef}) it follows, to leading order in $g$,
that
\bq \e_\tau=\frac{1}{\w\sqrt{2}}(3\pi g)^{1/3}.
 \eq
Now, using (\ref{ploon}), substituting (\ref{erg}) into (\ref{delta}),
and using (\ref{madef}),
we find that
\bq \Delta S[\z]=\w g^{-1/6}e^{-\frac{\pi}{2g}}\int dt_1
    \bl e^{-S_I^{(+)}[\z;\tau_0,t_1]}
     +e^{-S_I^{(-)}[\z;\tau_0,t_1]}\br.
 \label{zelta}\eq
Equation (\ref{zelta}) is
a significant result.
Concisely, it is the induced change in the canonical collective field
theory which results from the systematic inclusion of instanton
effects.   A lengthy analysis allows us to calculate from Eq.(\ref{zelta}) the
induced action as an integral over a local density.
Skipping a lot of details we simply state the results
\brr S_I^{(+)} &=& \frac{1}{\w}h_{00}(\z_-^{'}+\z_+^{'})
              +\frac{1}{\w^2}h_{01}(\z_-^{''}-\z_+^{''})
              +\frac{1}{\w^2}h_{10}(\dot{\z}_-^{'}-\dot{\z}_+^{'})
              +\frac{1}{\w^3}h_{11}(\dot{\z}_-^{''}+\dot{\z}_+^{''})
              +\cdots \nonumber \\
     S_I^{(-)} &=& \frac{1}{\w}h_{00}(\z_-^{'}+\z_+^{'})
              -\frac{1}{\w^2}h_{01}(\z_-^{''}-\z_+^{''})
              -\frac{1}{\w^2}h_{10}(\dot{\z}_-^{'}-\dot{\z}_+^{'})
              +\frac{1}{\w^3}h_{11}(\dot{\z}_-^{''}+\dot{\z}_+^{''})
              +\cdots.\nonumber \\
 \label{sinl}\err
where
\bq h_{mn}=\frac{\w^{m+n+1}}{m!n!}
    \int_{-\frac{\pi}{2}}^{\frac{\pi}{2}}dt
    \int_{-\infty}^{-\e_\tau} d\tau
    \J(\tau,t)\tau^mt^n,
 \label{hleft}\eq
\bq \z_\pm\equiv\z(\tau_0\pm\frac{\s}{2},t_1)
\label{not}\eq
and
\bq \J(\tau-\tau_0+\frac{\s}{2},t-t_1)=
    \frac{1}{(x(\tau-\tau_0+\frac{\s}{2})-\l_0^{(\pm)}(t-t_1))^2}-
\frac{1}{(x(\tau-\tau_0+\frac{\s}{2})-\l_{\emptyset}^{(\pm)}(t-t_1))^2}
 \eq
It is straightforward to compute the coefficients $h_{mn}$. We find, for
instance,
to leading order in $g$, that
\brr h_{00} &=& -\frac{4\sqrt{2}}{9} \nonumber \\
     h_{10} &=& -(\frac{8\pi g}{9})^{1/3} \nonumber \\
     h_{01} &=& -\frac{\pi\sqrt{2}}{9}.
 \label{hee}\err
In general, the $h_{mn}$ are found to have the following $g$
dependence,
\bq h_{mn}\sim\left\{
    \begin{array}{ll}
    g^{m/3} & ;\hspace{.2in} m\le 3 \\
    g & ;\hspace{.2in} m>3 \end{array}\right.
 \label{gee}\eq
Note, from (\ref{sinl}) and (\ref{gee}),
that, as the first index of $h_{mn}$ increases, that the corresponding
terms in $S_I^{(\pm)}$ depend on higher powers of
$g$. However, none of $h_{0n}$ have $g$ dependence for any value of
$n$.
We proceed to analyze the relative impact of these terms on generic
$N$-point
functions.  By putting (\ref{sinl}) back into (\ref{zelta})
we can find all relevant interaction vertices.  These are obtained by
Taylor expanding the exponentials in (\ref{zelta}).  For instance, we
obtain
the quadratic vertices $\frac{1}{\w^2}h_{00}^2\z_-^{'}\z_-^{'}$
and $\frac{1}{\w^3}h_{00}h_{10}\z_-^{'}\z_-^{''}$ where, as discussed
above, $h_{00}\sim 1$ and $h_{10}\sim g^{1/3}$.
It is clear that the effect of the second vertex, containing
$h_{00}h_{10}$,
on any $N$-point function, is suppressed by a factor
$g^{1/3} p/w$, where $p$ is a characteristic momentum,
when compared with effects arising solely from the
first vertex containing $h_{00}^2$.
This is true at tree level. At the quantum level,
 there may be some subtleties to this argument which
we will not discuss.  Similar considerations
apply to all other induced operators, involving higher $h_{mn}$.
It can thus be shown, provided
\bq p\lapp\w,
 \label{pcon}\eq
that, when working to leading order in $g$, we can
consistently drop all but the $h_{0n}$ terms in (\ref{sinl}).
Now, of the terms which remain, as $n$ increases, the corresponding
terms in $S_I^{(\pm)}$ depend on higher derivatives of $\z$.
Thus, the effect of any vertex, containing $h_{0n}$, on any
$N$-point function, is suppressed by a factor $(p/\w)^n$, relative to
effects arising from vertices containing only $h_{00}$.
If we further restrict momenta, such that
\bq p<<\w,
 \label{rees}\eq
we can then consistently neglect all but the $h_{00}$ terms
in (\ref{sinl}).  This results in a vast simplification of the
final result, so we will assume this approximation.  It would be
completely straightforward, however, to lift the restriction
(\ref{rees}),
and only require (\ref{pcon}).  One would then have to keep all
$h_{0n}$ terms in (\ref{sinl}).
It follows from (\ref{sinl}), that, to leading order
in $g$,
\bq \D S[\z]=2\w g^{-1/6}e^{-\frac{\pi}{2 g}}\int dt_1
    \exp\bl\frac{4\sqrt{2}}{3\w}\bpl\z'(\tau_0+\frac{\s}{2},t_1)
    +\z'(\tau_0-\frac{\s}{2},t_1)\bpr\br.
 \label{nl}\eq
Note however that equation (\ref{nl}) includes nonlocal interactions,
since it involves contributions coming from
$\z'$ evaluated simultaneously at $\tau_0-\frac{\s}{2}$ and also at
$\tau_0+\frac{\s}{2}$.  This is not surprising though, since we
have arrived at this result by integrating over single eigenvalue
instantons, which link effects on the left-hand side of the low-density
region with effects on the the right-hand side of this region, and
because there
is a finite separation between these two sectors. One may wish
to find some further approximation which would render the effective
theory
local.  This can be done as follows.  Provided we
consider momenta which satisfy (\ref{rees}), and provided also that
$\w\lapp\frac{1}{\s}$, the effective width of the low density region
as seen by any field will be essentially zero.  We therefore Taylor
expand $\z'(\tau_0\pm\frac{\s}{2},t_1)$ around the point
$(\tau_0,t_1)$,
thereby
taking
\bq \frac{1}{\w}\z'(\tau_0\pm\frac{\s}{2},t_1)
    =\frac{1}{\w}\z'(\tau_0,t_1)\pm\frac{\s\w}{2\w}\z''(\tau_0,t_1)
    +\cdots.
 \eq
Then, in a manner identical to the previous discussion,
we find that the
contributions coming from vertices which involve $\s$ are always
suppressed
by $(\s\w)p/\w$, where $p$ is a characteristic momentum.
Note that, since we now assume
$\w\lapp\frac{1}{\s}$, the factor $(\s\w)$ is $\lapp\O(1)$.
So, provided that
\bq p<<\w\lapp\frac{1}{\s},
 \eq
we may write the lowest order instanton-induced change
in the collective field action approximately, in local form, as
follows,
\bq \D S[\z]=2\w g^{-1/6}e^{-\frac{\pi}{2g}}\int dt
     e^{-\frac{2\sqrt{2}}{3\w}\z'(\tau_0,t)}.
 \eq
We have dropped the subscript $``1"$ on $t_1$ because it is now
superfluous.  This result can be written as a two-dimensional
integral over a density $\D S=\int dt d\tau \D\L$, where
\bq \D\L=2\w g^{-1/6}e^{-\frac{\pi}{2g}}
    \delta(\tau-\tau_0)e^{-\frac{2\sqrt{2}}{3\w}\z'(\tau,t)}.
 \eq
This is the final result of our calculation.

\end{document}